# Accelerating Correlation Power Analysis Using Graphics Processing Units (GPUs)

Hasindu Gamaarachchi, Roshan Ragel
Department of Computer Engineering
University of Peradeniya
Peradeniya, Sri Lanka
hasindu2008@gmail.com, roshanr@pdn.ac.lk

Darshana Jayasinghe
School of Computer Science and Engineering
University of New South Wales
Sydney, Australia
darshanaj@cse.unsw.edu

*Abstract*— Correlation Power Analysis (CPA) is a type of power analysis based side channel attack that can be used to derive the secret key of encryption algorithms including DES (Data Encryption Standard) and AES (Advanced Encryption Standard). A typical CPA attack on unprotected AES is performed by analysing a few thousand power traces that requires about an hour of computational time on a general purpose CPU. Due to the severity of this situation, a large number of researchers work on countermeasures to such attacks. Verifying that a proposed countermeasure works well requires performing the CPA attack on about 1.5 million power traces. Such processing, even for a single attempt of verification on commodity hardware would run for several days making the verification process infeasible. Modern Graphics Processing Units (GPUs) have support for thousands of light weight threads, making them ideal for parallelizable algorithms like CPA. While the cost of a GPU being lesser than a high performance multicore server, still the GPU performance for this algorithm is many folds better than that of a multicore server. We present an algorithm and its implementation on GPU for CPA on 128-bit AES that is capable of executing 1300x faster than that on a single threaded CPU and more than 60x faster than that on a 32 threaded multicore server. We show that an attack that would take hours on the multicore server would take even less than a minute on a much cost effective GPU.

## I. Introduction

Correlation Power Analysis (CPA) [1] is a type of power analysis based side channel attack that is used by adversary to break the secret key of an encryption system. CPA is made possible due to the correlation between the power consumption of the target device (measured as power traces) and the computation performed by the device during encryption or decryption. A CPA attack on an unprotected target device analyses about a few thousands of power traces and therefore is performed within a few minutes using commodity hardware. A vulnerability of such magnitude needs attention and therefore a large number of researchers work on coming up with countermeasures to protect against CPA attacks. Such countermeasures are usually verified against about 1.5 million power traces to prove their effectiveness. Such number of power traces are used as it is typically larger than the lifetime of the secret key in most practical systems [2].

On a general purpose computer with a single CPU, a typical verification against CPA attack takes about 40 minutes for 2000 traces. Therefore, assuming a linear variation in time, processing 1.5 million traces will take about 21 days. Even a high end server with 32 hardware threads would require about 18 hours for performing such a single verification. Therefore, those who implement countermeasures against CPA will have to spend a lot of time and resources to test their implementation, as they would try their countermeasure (and its variations) on a system several times. Therefore, a faster, efficient and cost effective method is definitely a requirement.

The present Graphics Processing Units (GPUs), having support for thousands of threads are ideal for a task like CPA, which has a large degree of parallelism. In addition, the GPU manufacturers now provide platforms and programming models to make the utilization of GPUs for general purpose calculations possible and easy. Therefore, GPUs are a cost effective method for speeding up CPA, yet providing even better performance benefits than a high performance server.

We use CUDA [3] (Compute Unified Device Architecture), which is the parallel programming model and platform provided by NVIDIA to be used in its GPUs for general purpose calculations to demonstrate the speed up possible for CPA using GPUs. For the mentioned 1.5 million power traces, the CUDA implementation on an NVIDIA Tesla GPU (C2075) takes less than 40 minutes. On modern GPUs this will take even lesser time.

The remaining part of the paper is organized as follows: Section II contains the related work and in Section III we present the background. Our methodology and implementation details are presented in Section IV. Our experimental setup and the results are presented in Section V. Finally we conclude the paper in Section VI.

## II. Related Work

To the best of our knowledge, there is currently no implementation for CPA on GPUs except [4]. However, there are several works that focus on just correlation analysis (not related to power analysis attack). Even though they are related in correlation analysis part, CPA includes several other steps which are again parallelizable.

Swamy et al. [4] have implemented the CPA algorithm on GPU. But they use CPU for finding correlation values while GPU is used only for finding hamming distance and maximum power values, stating that CPU is better for finding correlation coefficients. In our approach we assigned an individual CUDA thread for each key byte, for each key guess and for each bunch of power traces, achieving a higher data parallelism.

Kijsipongse et al. [5] have developed an algorithm to accelerate Pearson Correlation Matrix using CUDA. Here computations such as the sums of samples are calculated redundantly. In our approach, we split the evaluation into few steps and at the first step we calculate the redundantly used values and store them to be later reused by other steps.

Gembris et al. [6] also have implemented an algorithm for correlation analysis. They use a 2D thread indexing model. In our implementation, we use a 3D thread indexing model to increase the degree of parallelism and therefore doing better than what Gembris et al. have achieved.

There are several work in the literature related to DPA (Differential Power Analysis) on CUDA. However, DPA has certain problems like "ghost peaks" that may give erroneous results [1]; hence today CPA is the one that is widely used.

A classical CPU based CPA attack will consume quite lot of time even on a multithreaded multicore server. Therefore, a faster and cost effective method will be very useful. This paper contributes by proposing a CUDA implementation of CPA attack on GPU that is more than thousand times faster than a classical CPU implementation.

## III. BACKGROUND

### A. Correlation Power Analysis (CPA)

Side channel attacks refer to a broad class of attacks performed on a crypto system by using data obtained by observing physical implementation of the system. Data can be collected using methods such as computational time, power monitoring and acoustics. Power analysis uses power consumption data of the system. Correlation power analysis (CPA) is a special branch in power analysis, which is newer and superior than methods such as differential power analysis [1]. In CPA, the power consumption of the crypto system is modelled to a selection function. Then by applying statistical correlation methods the secret keys are deduced.

Here we consider the application of CPA to break the key of a 128-bit Advanced Encryption Standard (AES) system. Such an application is required in verifying countermeasures that are proposed against CPA attacks. In such verifications, first, power traces are obtained for a number of distinct encryptions processes using an oscilloscope that captures and records power traces or a similar method. We will need either the plain texts or the cipher texts, depending on the encryption round we use for analysis. If the plain text is used as input, the analysis must be done for the first round of AES and if the cipher text is selected, the analysis must be done on last round. By correlation analysis, the round key is derived. Finally, using the round key, the actual key can be derived. We consider the correlation analysis on the last round of encryption by using power traces and the cipher text as inputs.

Consider a situation where there are N power traces and N corresponding cipher texts. Each power trace has M number of sampling points. The $j^{th}$ sample point of the $i^{th}$ power trace will be written as $W_{i,j}$. The selection function for the $i^{th}$ cipher text with respect to the appropriate sub key will be written as $H_i$. An estimate for the correlation for $j^{th}$ sampling point for a certain sub key can be found using Equation (1) [1].

$$\hat{\rho} = \frac{N \sum_{i=0}^{N} W_{i,j} H_i - \sum_{i=0}^{N} W_{i,j} \sum_{i=0}^{N} H_i}{\sqrt{N \sum_{i=0}^{N} W_{i,j}^2 - (\sum_{i=0}^{N} W_{i,j})^2} \sqrt{N \sum_{i=0}^{N} H_i^2 - (\sum_{i=0}^{N} H_i)^2}} \quad (1)$$

The CPA process can be divided into the following phases.

**Phase1 - Finding power model statistics**
The AES key is divided into sub keys of 8 bits (1 byte). Then for all possible sub keys and byte positions, the value of the selection function $H_i$ with regard to each and every cipher text is calculated. Then, the statistical sums of the selection function per each (sub key, key position) combination, $\sum_{i=0}^{N} H_i$ and $\sum_{i=0}^{N} H_i^2$ are found in this phase.

**Phase2 - Finding power trace statistics**
For each combination of sub key, byte position and sample point of the power traces, the statistical sums for power trace values $\sum_{i=0}^{N} W_{i,j}$ and $\sum_{i=0}^{N} W_{i,j}^2$ are calculated. Also $\sum_{i=0}^{N} W_{i,j} H_i$ must be found using $H_i$ that was calculated in Phase1.

**Phase3 - Finding maximum correlation**
For each combination (of sub key, byte position), the maximum correlation out of all wave sample points is found.

**Phase4 - Deriving the round key**
Sub keys that have maximum correlation is selected.

### B. Compute Unified Device Architecture (CUDA)

CUDA is a general purpose parallel programming model and platform introduced by NVIDIA to facilitate the usage of its GPUs for efficient computations of complex calculations [3]. Graphic processing naturally demands great amount of parallelism, hence they have support for a large number of cores and threads. While typical modern general purpose processors have less than 10 cores, current GPUs consists of hundreds of cores and supports thousands of threads providing a large amount of simultaneous parallel executions.

GPUs are very efficient for applications where there exists a large amount of data level parallelism. That is, the same program is independently executed on a large amount of data items in parallel with relatively more arithmetic operations compared to memory accesses. Since GPUs are devoted for calculations rather than caching and flow control, programs that have less divergent flow control and relatively less number of memory accesses can execute extremely fast on GPUs compared to CPUs.

The CUDA toolkit provided by NVIDIA enables using C programming language to manipulate the GPU at a high level. It allows a special type of functions called kernels, each one of them is defined for a single thread (like template) but executes on multiple threads when called. CUDA C also provides an Application Programing Interface (API) for tasks like allocating memory in the device, copying data to and from the device. It also provides a thread model that supports up to 3 dimensions of thread indexing, which provides the programmer convenience to naturally map vectors, matrices or volumes as necessary.

CUDA devices have several types of memories, such as global memory, shared memory, local memory, and registers. Each one has its special properties, where the selection of the most suitable memories will yield a higher performance.

When maximizing performance, factors like thread block size and numbers, register usage, memory coalescing, usage of caches, memory alignment must also be considered.

## IV. CPA WITH CUDA

In Section III, it is clearly shown that the CPA algorithm matches exactly what CUDA can most efficiently compute. In this section, we describe the methodology and the implementation of CPA on a GPU using CUDA.

An intuitive way of parallelizing the CPA algorithm, which we implemented first, is to just map each combination of sub key and byte position into separate threads to form single CUDA kernel with 2D thread model. It had significant amount of performance improvement compared to a CPU implementation. However, in this implementation, a single thread will have to look after all the calculation per each sample point. Such a mapping introduces nested loops to a thread while increasing the register consumption per thread and therefore reduces the amount of simultaneous executions.

Therefore, we separated operations into phases as we described in Section III based on the amount of parallelism possible and implemented the phases in separate kernels while mapping them to the most suitable thread indexing models and data structures to achieve the best performance possible. The comparison of performance between the two approaches (the intuitive one verses the one described in this paragraph) is presented in Section V. In addition to the new mapping, we used many other strategies to further enhance the performance, which we describe one by one below.

### A. Thread models based on the degree of parallelism

Different phases described in Section III have different levels of parallelism. To maximize the performance, identification of such is critical, so is the matching them with proper thread dimensions.

Consider Fig. 1, which depicts the thread model we used for Phase1. Vertical axis depicts the possible values that a sub key can take. A sub key is 8 bits long and therefore the possible values change from 0 to 255. An AES key is 128 bits long and therefore it has 16 sub keys, which are denoted by the horizontal axis. Therefore, each square denotes a unique sub key and position combination and hence the grid covers all possible keys. Per each combination, the selection function statistics (marked per thread number (p, q) by using an arrow in Fig. 1) must be calculated independently. This means that each combination can be handled in parallel, hence we assign separate threads per each square as shown in Fig. 1.

Now consider Fig. 2, which depicts the thread indexing model for Phase2. This phase is the most computationally intensive phase (the time consumed by each phases are given in Section V). Therefore, the thread indexing model for this phase is the most significant one.

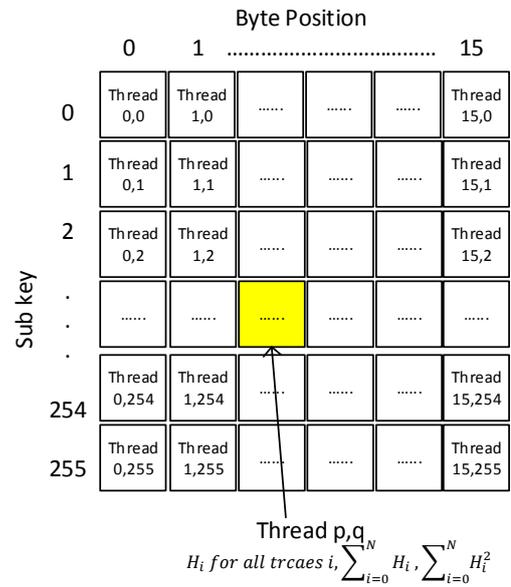

Fig. 1. Thread structure for Phase1

In Fig. 2, the horizontal and vertical axes are the same as in Fig. 1. However, Fig. 2 has a third axis, which denotes each sample point of a trace. The value for "m" in Fig. 2, that is the index of the final sample depend on the sampling rate of the equipment and generally will be in the order of a few ten thousands. Now, each cube in the 3D model represents a unique combination of sub key, byte position and sample point. The 3D mesh covers all possibilities. Per each combination, the wave statistics as shown per thread (p,q,r) must be calculated. A separate thread can be assigned to each cube, as the calculation per a cube is independent to the other cubes. However, the calculation for a single cube cannot be further parallelized as they are all sums, which therefore must be serially executed.

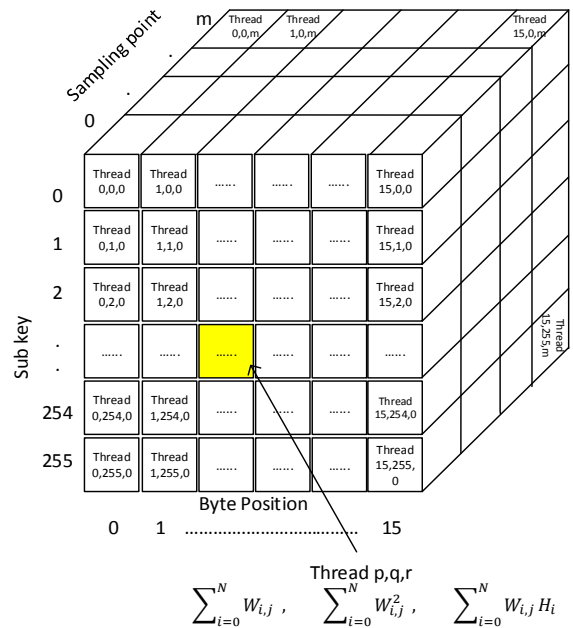

Fig. 2. Thread structure for Phase2

Usage of this 3D thread indexing model is a critical factor that improves the performance of our implementation compared to many others. We claim this as one of the main contributions that stands out in our correlation analysis implementation compared to other GPU implementations of correlation analysis.

For Phase3, the thread model is exactly similar to the one presented in Fig. 1 except that now the calculation performed per each thread is finding of the maximum correlation instead of calculating selection function statistics. The axes remain the same so the squares. Therefore, we use a 2D thread model. The thread (p,q) now has to calculate the correlation for each sample point and to get the maximum correlation out of them.

Phase4 selects the sub key with the maximum correlation for each bye position. This can be mapped to a 1D thread model. However, since the number of parallel jobs is very limited here, calling a CUDA kernel is expensive and therefore we perform this phase in the host CPU. When compared to other phases, Phase4's computational time is negligible.

### B. Selection of the block sizes

The memory latency in CUDA is hidden by the amount of computation performed in parallel. Occupancy of the multiprocessors or the ratio of the number of active warps to the number of maximum supported warps is an important parameter that will determine the performance [7]. The more occupancy we have, the more the memory latency is hidden hence improving performance. The occupancy depends on the number of registers used by a thread and also the block size. Therefore, after finding the number of registers necessary per thread, CUDA occupancy calculator can be used for finding a suitable block size. Also for reasons like providing best memory coalescing and preventing wastage of warps, the block dimensions being a values like 16 or 32 is more preferable. Therefore, suitable block sizes that best suit all these factors were calculated. Table I depicts the block dimensions we used.

### C. Coalesced memory access

In CUDA, when parallel threads have to access different memory locations in the global memory, if each access is performed one by one, then the accesses will be serial, causing decrease in performance. Therefore, what is done is a set of memory accesses are collected together and they are read as a chunk. This is called coalesced memory access. The cache in CUDA is used for the purpose of coalescing memory accesses but not for usual spatial and temporal locality as in CPU caches. For this phenomenon to take place, the memory accesses by contiguous threads must be contiguous. If memory accesses of contiguous threads are distributed, then more memory accesses will occur, declining performance. In our program, we use the CUDA API function to allocate memory and therefore only one dimensional arrays can be allocated in this fashion.

TABLE I. CUDA KERNELS AND THE SELECTED BLOCK SIZES

| Kernel | Block Dimension |
|---|---|
| Phase 1 (Power model statistics) | 16,16 |
| Phase 2 (power trace statistics) | 16,16 |
| Phase 3 (maximum correlation) | 16,16,1 |

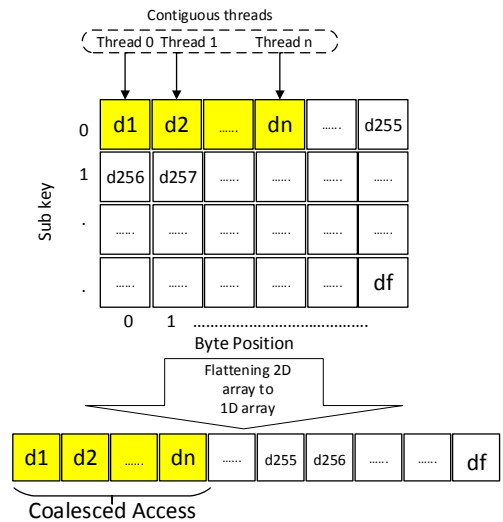

Fig. 3. Flattening a 2D array to a 1D array in row major format to support coalesced memory access based on the thread structure

Multidimensional arrays must be flattened to one dimensional arrays by either in row major or column major fashion. However, a proper method must be selected for each array based on the access pattern by the contiguous threads in order to minimize serialization of memory accesses.

To explain the strategy adapted, we use the array that is used to store the sum of selection function values. The upper part of Fig. 3 depicts this array. The array is used to store the sum of selection function values per each sub key, byte position combinations. The rows in the array correspond to sub keys and columns to byte positions. In each slot, we keep our data, which are marked as d1, d2, d3 and so on. Therefore, naturally this array must be a 2D array.

Phase3 of our algorithm accesses this array to read data. Note that the axes in Fig. 1 and Fig. 3 are the same. We explained this in Phase3 that it uses a thread model similar to that of Fig. 1. Therefore, each thread accesses the corresponding slot in the array. Say threads (0,1), (0,2), (0,3) …. (0,n) are launched in parallel by the scheduler. Those threads will access the highlighted data in the upper part of Fig. 3. The lower part of Fig. 3 shows how the array is flattened to a linear array using row major to actually store the data in CUDA memory. Note that the highlighted data, which correspond to a simultaneous access by the parallel threads are still contiguous. The memory accesses by the running threads are therefore coalesced. If the array was flattened in column major d1, d256…. will be the contiguous ones. Therefore, the memory access would have been scattered thus reducing the effective usage of cache, causing the memory bus to saturate.

## V. EXPERIMENTAL SETUP AND RESULTS

### A. Equipment, tools and configurations

For the experiments, we used an NVIDIA Tesla C2075 graphics card which has CUDA compute capability 2.0 with 448 CUDA cores and 6GB of memory. The card is fixed on a general purpose computer with an i3 530 CPU of 3.1 GHz, with 6GB of RAM. CUDA 5.0 toolkit was used for implementation and the code is entirely written in CUDA C.

To compare the performance of GPU with serial execution on CPU, we implemented the same algorithm in C in single threaded fashion and tested on the CPU of the host computer. We also compared the performance of our CUDA implementation with a parallel CPU execution on a multithreaded multicore CPU. For that, we implemented the same algorithm in C using pthreads library and tested it on a HP ProLiant DL380p Gen8 high performance server consisting 32 CPU threads (16 cores of Xeon-E5-2670) and 256GB of RAM. Finally, we also tested with a few other GeForce GPUs, which are usually used for gaming purposes.

In all experiments, we assumed that all necessary data are residing in the CPU RAM. The time measurements for serial CPU implementation only consists of time spent for function calls and calculations of phases 1, 2, and 3 in the algorithm we described in Section III. The time measurements in CUDA implementation consists of function calls and calculations of phases 1,2,3 as well as time overhead to copy data from the RAM to device (GPU) memory and vice versa.

The inputs for the programs are the cipher text and power traces. We use two sets of power traces, one having 2000 traces with 12000 sampling points (dataset1) and the other having 9000 traces with 48000 sampling points (dataset2) and their corresponding cipher texts. Dataset1 was obtained from SASEBO Evaluation Board website [8] and dataset2 was obtained from our own experiments in a SASEBO board [9].

*B. Comparison of performance on various devices*

First, we compared the performance of the intuitive implementation with the most optimal method we implemented as described at the beginning of Section IV on Tesla C2075. The intuitive way of implementation on CUDA took 8.39 seconds while the optimized method on CUDA took only 3.31 seconds giving a performance improvement of about 2.5 times on dataset1. Therefore, in the following paragraphs we use that most optimal implementation to compare performance on CUDA with respect to CPU.

Fig. 4 shows the comparison of time consumption of various devices to execute CPA algorithm on dataset1. Since the time for single threaded CPU is extremely large, the comparison between other devices will not be visible on the graph if it is included. Therefore, we have included the graph for all devices except the single threaded CPU. Clearly Tesla C2075 which is a device specially designed for high performance computing has the highest performance of all. GT-540M and GT-720M are GPUs found on laptops for gaming and other home usage graphic needs. However, they have better performance than the high performance server (noted as 32 threaded CPU) by considerable amount of time.

GTX-770 is a recent gaming graphics card that has compute capability 3.0. The performance of that card is closer to the Tesla C2075 high performance graphics card. Therefore, on a latest device like Tesla K40, which has compute capability 3.5, the performance improvement will be much higher than the numbers reported for Tesla C2075 here.

Based on our experiments, we also noticed that the increase of speedup with data between a single threaded CPU and Tesla C2075 implementation increases better than linearly.

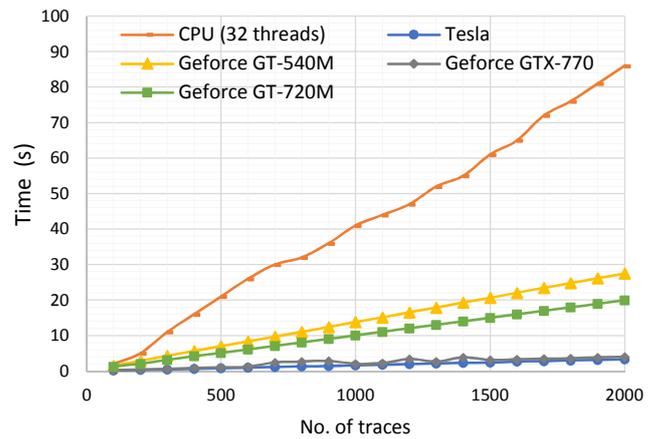

Fig. 4. Execution time of dataset 1 on different devices

Speedup is defined as the time elapsed on the single threaded CPU divided by time on the Tesla card. To show how much of gain can be attained, we tabulated the time for 1000 traces in dataset2 on the tesla card and the single threaded CPU in Table II. As shown in the table, a speedup of about 1300 times was achieved. If the data size is further increased this gain would be even higher. The reason is, the time taken to copy data between RAM and CUDA memory and launching kernels. For larger data sets, the copying time and kernel launch times are smaller when compared to the computational time. Therefore, we see a larger gain when the data set is larger. Apart from this, factors such as cache hits, loop divergences, memory coalescing and external factors like CPU usage by operating system will effects the speedup.

*C. GPU vs. multi-threaded multicore server*

Among the CUDA devices, the best device we had access to be was a Tesla C2075 and from the CPU domain it was the 32 threaded server. Therefore, we used a larger dataset of traces containing 48000 sample points (dataset2) to compare CUDA and multithreaded server for which the results are given in Fig.5.

TABLE II. THE SPEEDUP ON TESLA C2075 WITH RESPECT TO THE SINGLE THREADED CPU

| Time on CPU (s) | Time in CUDA (s) | Speedup |
|---|---|---|
| 8856 | 6.57 | 1348 |

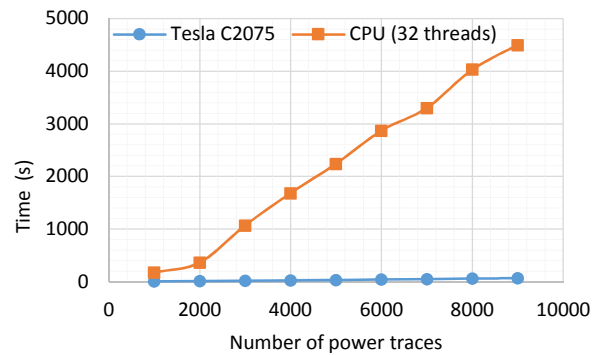

Fig. 5. Variation of elapsed time with the data set size for Tesla C2075 and 32 threaded CPU

Despite the cost of the multicore server compared to the Tesla card, the CUDA implementation is efficient for CPA. When the data size is larger, the CUDA implementation still gives a speedup of about 66 times the performance on the high performance server. For example, consider the execution time for 8000 traces. The high performance server takes about 4000 seconds which is more than an hour. However Tesla C2075 takes about 60 seconds. Therefore, imagine a situation of several million traces. The high performance server will take weeks to complete, but C2075 will complete in even less than an hour.

### D. Effects of single precision and double precision numbers

All experiments presented up to now are performed using double precision floating point numbers as data values. Table III compares the effect of using single precision and double precision numbers on both GPU and CPU. Here we just have compared the CUDA implementation with a single threaded program on a CPU for 100 traces with 48000 sample points.

Even though the CPU time does not change much, when single precision floating point representation is used instead of double precision, in GPU using of single precision floating points increases the performance by considerable amount. In the 64 bit CPU used, whether we use single or double precision, a value resides on a single register. Hence the time taken on CPU does not change based on this factor. However, in CUDA implementation, double precision floats use 2 registers while single precision floats use just one register. Therefore, the floating point representation will affect the efficiency in CUDA by a considerable amount. In devices with CUDA compute capability 3.0, the performance of double precision floating points have been further enhanced now. Therefore, this shows that in modern and future devices CUDA implementation would consume even lesser time.

### E. The time consumption distribution for various kernels of the CUDA program

Table IV contains the summary of the results obtained by running the Nsight visual profiler [10] on our CUDA implementation. A small dataset of 100 traces with 48000 sampling points have been used for the profiling, as a larger datasets needs long time for profiling.

The phase1 kernel takes negligible amount of time compared to the others. Therefore, as per Amdahl's law, optimizing it further will not have any significant impact on the overall performance. The kernel that takes the most amount of time is phase2. However, the performance limiter in phase2 is the computation. This may be optimized further by scheduling the instructions to utilize the pipeline as much as possible. However, this is mostly a hardware limitation. In most recent GPU devices with Kepler architecture comprising compute capability 3.0, CUDA cores are much more powerful. Therefore, it is quite clear that on modern devices the time consumed will be even lesser than what we have presented in this section.

TABLE III. EFFECT OF FLOATING POINT TYPE ON PERFORMANCE

|  | Time on CPU (s) | Time in Tesla C2075 (s) |
|---|---|---|
| Double precision | 299.2 | 0.67 |
| Single precision | 296.3 | 0.49 |

TABLE IV. STATISTICS ABOUT CUDA KERNELS

| CUDA kernel | Consumed time as a percentage | Performance Limiter |
|---|---|---|
| Phase1 (Power model statistics) | 00.0% | Instruction and Memory Latency |
| Phase2 (power trace statistics) | 85.8% | Computation |
| Phase3 (maximum correlation) | 14.2% | Computation |

## VI. CONCLUSIONS

In this paper, we have shown that the CPA algorithm on a GPU can execute more than 1300x faster than a generic single threaded CPU. When compared to a multithreaded version run on a multicore high performance server, the GPU implementation is about 60x faster. We further have shown that, on the latest NVIDIA GPUs with compute capability 3.5 this would be even faster. Such improvements will help researchers who work on countermeasures for CPA attacks to verify their research faster and therefore work on their research effectively. Using GPUs for CPA is a cost effect method, which takes practically feasible time.

In future, we plan to use new features of latest Kepler architecture with compute capability 3.5 and the latest CUDA tool kit 6.0 to explore the possible performance improvements. We will try out new features like dynamic parallelism and unified memory access. We also plan to extend our implementation to make use to multiple GPUs simultaneously to further reduce time for massive datasets.